\documentclass[twocolumn,secnumarabic,amssymb, nobibnotes, aps, prl, superscriptaddress]{revtex4-1}
\usepackage{graphicx}                    
\usepackage{amsmath}
\usepackage{amssymb}                    
\usepackage{color}                       
\usepackage{subfigure}
\usepackage{url}

\begin{document}

\title{Alignment and scaling of large-scale fluctuations in the solar wind}%

\author{R. T. Wicks}
\email{robert.t.wicks@nasa.gov}
\affiliation{NASA Postdoctoral Program Fellow, Goddard Space Flight Center, Greenbelt, MD, USA}
\author{A. Mallet}%
\affiliation{Rudolf Peierls Centre for Theoretical Physics, University of Oxford, Oxford, OX1 3NP, UK}
\author{T. S. Horbury}
\affiliation{Space and Atmospheric Physics Group, Imperial College London, London, SW7 2AZ, UK}
\author{C. H. K. Chen}
\affiliation{Space Sciences Laboratory, University of California, Berkeley, California 94720, USA}
\author{A. A. Schekochihin}%
\affiliation{Rudolf Peierls Centre for Theoretical Physics, University of Oxford, Oxford, OX1 3NP, UK}
\author{J. J. Mitchell}
\affiliation{Space and Atmospheric Physics Group, Imperial College London, London, SW7 2AZ, UK}

\date{\today}
\begin{abstract}
We investigate the dependence of solar wind fluctuations measured by the Wind spacecraft on scale and on the degree of alignment between oppositely directed Elsasser fields. This alignment controls the strength of the non-linear interactions and, therefore, the turbulence. We find that at scales larger than the outer scale of the turbulence the Elsasser fluctuations become on average more anti-aligned as the outer scale is approached from above. Conditioning structure functions using the alignment angle reveals turbulent scaling of unaligned fluctuations at scales previously believed to lie outside the turbulent cascade in the `$1/f$ range'. We argue that the $1/f$ range contains a mixture of non-interacting anti-aligned population of Alfv\'{e}n waves and magnetic force-free structures plus a subdominant population of unaligned cascading turbulent fluctuations.
\end{abstract}
\maketitle
{\it Introduction.} The solar wind is a hot, tenuous plasma that flows away from the Sun at supersonic speeds. Turbulence transports energy from the driving `outer' scale to smaller scales via non-linear magnetohydrodynamic (MHD) interactions of magnetic $\textbf{B}$ and velocity $\textbf{V}$ fields, until kinetic effects and dissipation become important close to the ion gyroscale. In fast solar wind ($|V| > 600$ km/s), a `$1/f$' scaling of magnetic-field power spectra is observed at low spacecraft frequencies, $f$ \cite{Goldstein95, Horbury96, Wicks10}. Slowly evolving structures are advected at supersonic speeds past spacecraft, so the observed spacecraft frequency of a fluctuation is proportional to its characteristic wave number (scale) $k$ \citep{Taylor}. The energy spectrum in the $1/f$ range is, therefore, expected to scale as $E(k) \propto k^{-1}$. A steeper spectrum close to $k^{-5/3}$ associated with turbulence is observed at higher spacecraft frequencies in the `inertial range' and there is a spectral break between the two regimes \citep{Matthaeus86, Velli90, Horbury96, Wicks10}; at $1$ AU this typically occurs at $f \sim 10^{-3}$ Hz. Studies have shown \citep{Belcher71, Matthaeus86, Roberts87, Bavassano92, Horbury96, Matthaeus07} that the power spectral density of fluctuations in the low-frequency band decreases with distance from the Sun as $R^{-3}$, consistent with these scales containing non-interacting Alfv\'{e}n waves \citep{Hollweg90}, and thus these large-scale fluctuations are thought to have originated at the Sun and travelled outwards with relatively little {\it in-situ} modification. In this Letter, we argue that this interpretation is incomplete as nonlinear interactions occur at larger scales than previously thought (see also \cite{Verdini12}).
\par
Recently the concept of scale-dependent alignment has become prominent in theoretical and numerical studies of MHD turbulence \citep{Boldyrev06, Mason06, Perez09, Beresnyak08, Beresnyak09, Beresnyak11}. Scale-dependent alignment is the tendency for the angle between fluctuations of $\textbf{B}$ and $\textbf{V}$ in the plane perpendicular to the mean magnetic field $\textbf{B}_0$ to decrease with increasing $k$. 
Attempts to measure alignment in the inertial range of the solar wind produce no evidence of scaling \citep{Podesta09b, Hnat11}, although the ability to measure the scale dependence is limited by instrument noise characteristics. These studies do, however, find a scaling of the alignment in the $1/f$ range, which is unexpected given the previous interpretations of these large-scale fluctuations as non-turbulent.
\par
Here we study the alignment of Alfv\'{e}nic fluctuations in the $1/f$ range. We use Elsasser variables \citep{Elsasser, Tu89, Tu90} to characterize the Alfv\'{e}n waves that travel sunward and anti-sunward in the plasma frame. We define the angle $\phi$ between fluctuations in the Elsasser fields in the plane perpendicular to $\textbf{B}_0$ as the alignment angle. This angle is geometrically related to the alignment angle between fluctuations $\delta \textbf{B}$ and $\delta \textbf{V}$ but is not completely determined by it. The alignment angle $\phi$ is important because it controls the strength of the non-linear interaction \citep{DMV1980} and, as we are about to see, allows one to sort the large-scale fluctuations into steep-scaling `turbulent' and shallow-scaling `non-turbulent' populations. 
\par
{\it Data.} We use Wind spacecraft observations of solar wind magnetic field $\textbf{B}$, velocity $\textbf{V}$, and proton number density $n_p$ at cadence $\delta t = 3$~s made by the MFI and 3DP instruments during a 6-day-long fast stream interval observed between days 14 and 20 of 2008. The average solar wind conditions were: $|V| = 660$~km/s, $|B| = 4.4$~nT, $n_p = 2.4$~cm$^{-3}$, Alfv\'{e}n speed $V_A = 62$~km/s, and the ratio of thermal to magnetic pressure for protons $\beta_p = 1.2$. Similar fast streams recurred five times in succession due to a long-lived, low-latitude coronal hole. A further three of these fast streams were also analyzed and provide quantitatively similar results to those shown here; one stream was excluded because it coincided with a large data gap. The same analysis performed on Ulysses spacecraft data when in fast wind over the poles of the Sun also shows qualitatively the same results as described below. 
\par
\begin{figure}
\includegraphics[width=\columnwidth]{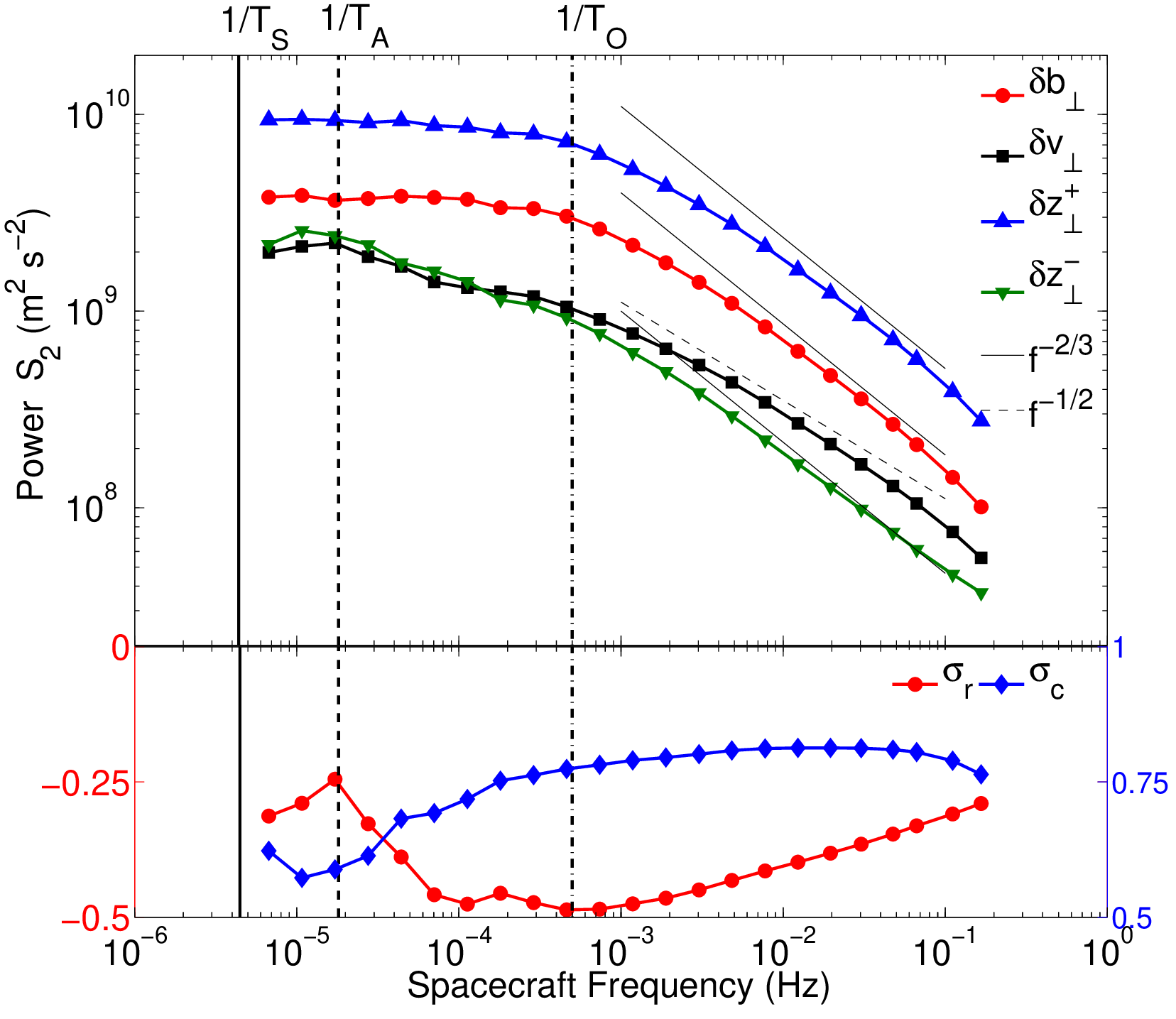}%
\caption{Structure functions of Wind data during a 7-day-long fast stream. $T_S$ is the estimated travel time of the solar wind from the Sun to the spacecraft, $T_A$ is the time it takes for the largest scale over which Alfv\'{e}n waves can interact to be advected past the spacecraft, and $T_O$ is the time scale corresponding to the outer scale of the turbulence, defined as the spectral break in the $\textbf{B}$ structure functions. The bottom panel shows normalized cross helicity $\sigma_c$ and residual energy $\sigma_r$ calculated from this data.}
\label{fig:1}
\end{figure}
We use the Alfv\'{e}n-normalized magnetic field $\tilde{\textbf{B}} = \textbf{B}/\sqrt{4\pi m_pn_p}$. Elsasser variables, $\textbf{Z}^{\pm} = \textbf{V} \pm \tilde{\textbf{B}}$, are re-defined so that $\textbf{Z}^+$ are anti-sunward and $\textbf{Z}^-$ sunward propagating fluctuations in the plasma frame. We are interested in alignment, so we use only the projection of the fluctuating fields on to the plane perpendicular to the local mean magnetic field $\textbf{B}_{0}$ at a time scale $\tau$:
\begin{align}
\textbf{B}_0(t,\tau) =& \frac{\delta t}{\tau}\sum\limits_{t' = t}^{t' = t+\tau}{\textbf{B}(t')},\label{eq:B0}\\
\delta\textbf{x}(t,\tau) =& \textbf{X}(t) - \textbf{X}(t+\tau),\label{eq:dx}\\
\delta\textbf{x}_{\perp}(t,\tau) =& \delta\textbf{x}(t,\tau) - \left(\delta \textbf{x}(t,\tau)\cdot\hat{\textbf{B}}_0(t,\tau)\right)\hat{\textbf{B}}_0(t,\tau),\label{eq:dxperp}
\end{align}
where $\hat{}$ denotes unit vectors and \textbf{X} can be $\tilde{\textbf{B}}$, $\textbf{V}$, $\textbf{Z}^+$ or $\textbf{Z}^-$. In the plots presented below, the time scale $\tau$ is converted into a frequency in the spacecraft frame to facilitate comparison with Fourier spectra: $f = 1/\tau$. A logarithmically spaced range of time scales $6$ s $< \tau < 2\times10^5$ s is used to investigate the inertial and $1/f$ ranges of the fast solar wind.
\par
{\it Structure functions.} In Fig.~\ref{fig:1}, we show the second-order structure functions of all four vector fields perpendicular to the magnetic field:
\begin{equation}
S_2(\delta\textbf{x}, \tau) = \frac{1}{N}\sum\limits_{t = t_1}^{t = t_2}{\left| \delta\textbf{x}_{\perp}(t,\tau) \right|^2} = \left< \left| \delta\textbf{x}_{\perp}(t,\tau) \right|^2 \right>,
\label{eq:S2}
\end{equation}
where $N$ is the number of samples in the time period $t_1 < t < t_2$. 
The scaling exponent of the structure functions $\alpha$, where $S_2(\delta\textbf{x}, \tau) \propto \tau^{-\alpha} \propto f^{\alpha}$, is related to the Fourier spectral index $\gamma$ by $\gamma = \alpha - 1$ \cite{MoninYaglom75, MarschTu97}.
\par
\begin{figure}
\includegraphics[width=\columnwidth]{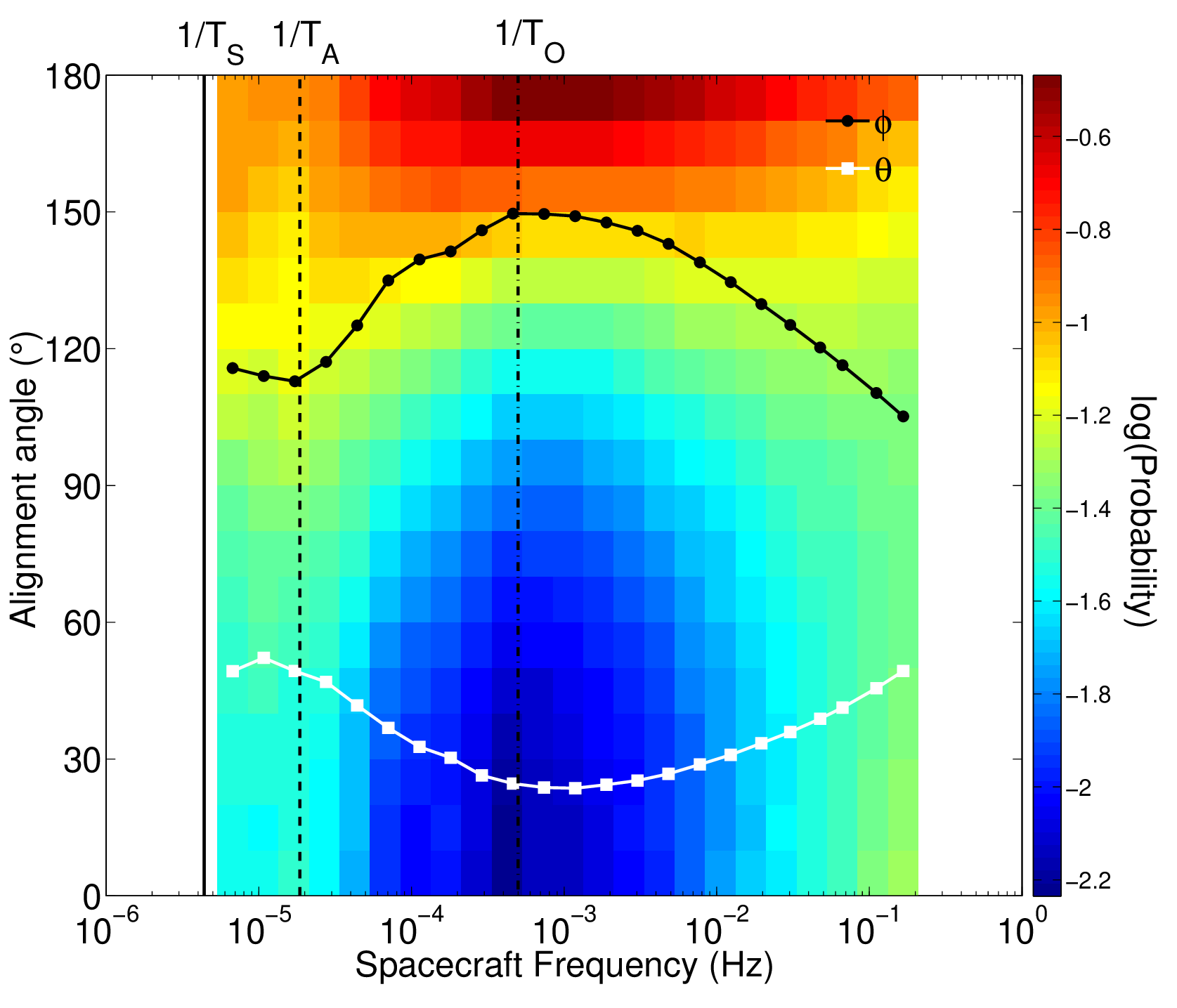}%
\caption{The mean (black circles) and the probability distribution (color scale) of $\phi$ at different scales. The angle $\theta$ between $\delta\textbf{v}_{\perp}$ and $\delta\textbf{b}_{\perp}$ is shown as white squares for comparison.}
\label{fig:2}
\end{figure}
The vertical lines in Fig.~\ref{fig:1} show important time scales for this period of solar wind. $T_S = 1$~AU$/|V| = 2.3\times10^5$~s is the approximate time the solar wind has taken to travel from the Sun to the Wind spacecraft. $T_O = 2\times10^3$~s is the approximate time scale associated with the outer scale, defined as the scale at which $S_2(\delta\textbf{b})$ rolls over from flat in Fig.~\ref{fig:1} ($\alpha = 0$ corresponding to the spectral index $\gamma = -1$, the $1/f$ range) to an inertial range scaling ($\alpha \approx -2/3$). $T_A = L/|V| = 5.4\times10^4$ s is the approximate time scale associated with the advection past the spacecraft of the largest separation $L$ two counter-propagating Alfv\'{e}n waves can have and still meet one another in the time the solar wind has taken to propagate from the Sun to the spacecraft. To calculate $T_A$ we estimate the dependence on heliocentric distance $R$ as follows: $|B| \propto R^{-1.5}$ and $\rho_i \propto R^{-2}$ and thus $V_A \propto R^{-0.5}$ and solve for the distance from the Sun $L$ that the slower of the two Alfv\'{e}n waves ($|V| + V_A$ and $|V| - V_A$) must start so that the faster wave just meets it at 1 AU. Thus the spacecraft frequencies between $f\sim1/T_S$ and $f\sim1/T_A$ represent spatial structure between different source regions in the corona, since they cannot have interacted during transit from the Sun. The range of frequencies between $1/T_A$ and $1/T_O$ contains fluctuations that may have interacted; on these scales, all structure functions are relatively flat, with $S_2(\delta\textbf{b}_{\perp},\tau)$ and $S_2(\delta\textbf{z}^+_{\perp},\tau) \propto f^0$. Frequencies higher than $1/T_O$ show all variables with scaling typical of turbulence in the fast solar wind: $S_2(\delta\textbf{b}_{\perp},\tau) \propto S_2(\delta\textbf{z}^+_{\perp},\tau) \propto S_2(\delta\textbf{z}^-_{\perp},\tau) \propto f^{-2/3}, S_2(\delta\textbf{v},\tau) \propto f^{-1/2}$ \cite{Roberts10, Boldyrev11}.
\par
{\it Alignment angle.} In order to investigate correlations between Elsasser fluctuations, we calculate the local scale-dependent $\phi$:
\begin{align}
\phi(t, \tau) &= \arccos{\left[\frac{\delta \textbf{z}^+_{\perp}(t, \tau) \cdot \delta \textbf{z}^-_{\perp}(t, \tau)}{\left| \delta \textbf{z}^+_{\perp}(t, \tau) \right| \left| \delta \textbf{z}^-_{\perp}(t, \tau) \right|}\right]}.
\end{align}
In Fig.~\ref{fig:2}, we show the mean (in black) and the probability distribution of $\phi$ at each scale. The distribution of $\phi$ has been discretized using $10^{\circ}$ wide bins. At all scales, the distribution of $\phi$ covers the full range of possible values and is peaked at $180^{\circ}$. The mean is not the most probable value at any scale and is strongly dependent on the tail of the distribution that extends towards $0^{\circ}$. The mean values of $\theta$, the angle between $\delta\textbf{v}_{\perp}$ and $\delta\textbf{b}_{\perp}$, calculated in a similar manner to $\phi$, are also shown (in white).
\par
The frequencies $f \sim 1/T_A$ and $f \sim 1/T_O$ both coincide with marked changes in behavior of the distribution and mean of $\phi$ in Fig.~\ref{fig:2}. The mean alignment angle, $\left<\phi\right>$, increases between $f\sim1/T_A$ and $f\sim1/T_O$, but then rolls over and decreases at higher frequencies, similar to previous observations of $\theta$ \cite{Podesta09b, Hnat11}. The peak in the distribution at $180^{\circ}$ grows as frequency increases in the range $1/T_A \lesssim f \lesssim 1/T_O$ but then flattens and begins to decrease where instrument noise becomes important, as discussed below. 
\par
\begin{figure}
\includegraphics[width=\columnwidth]{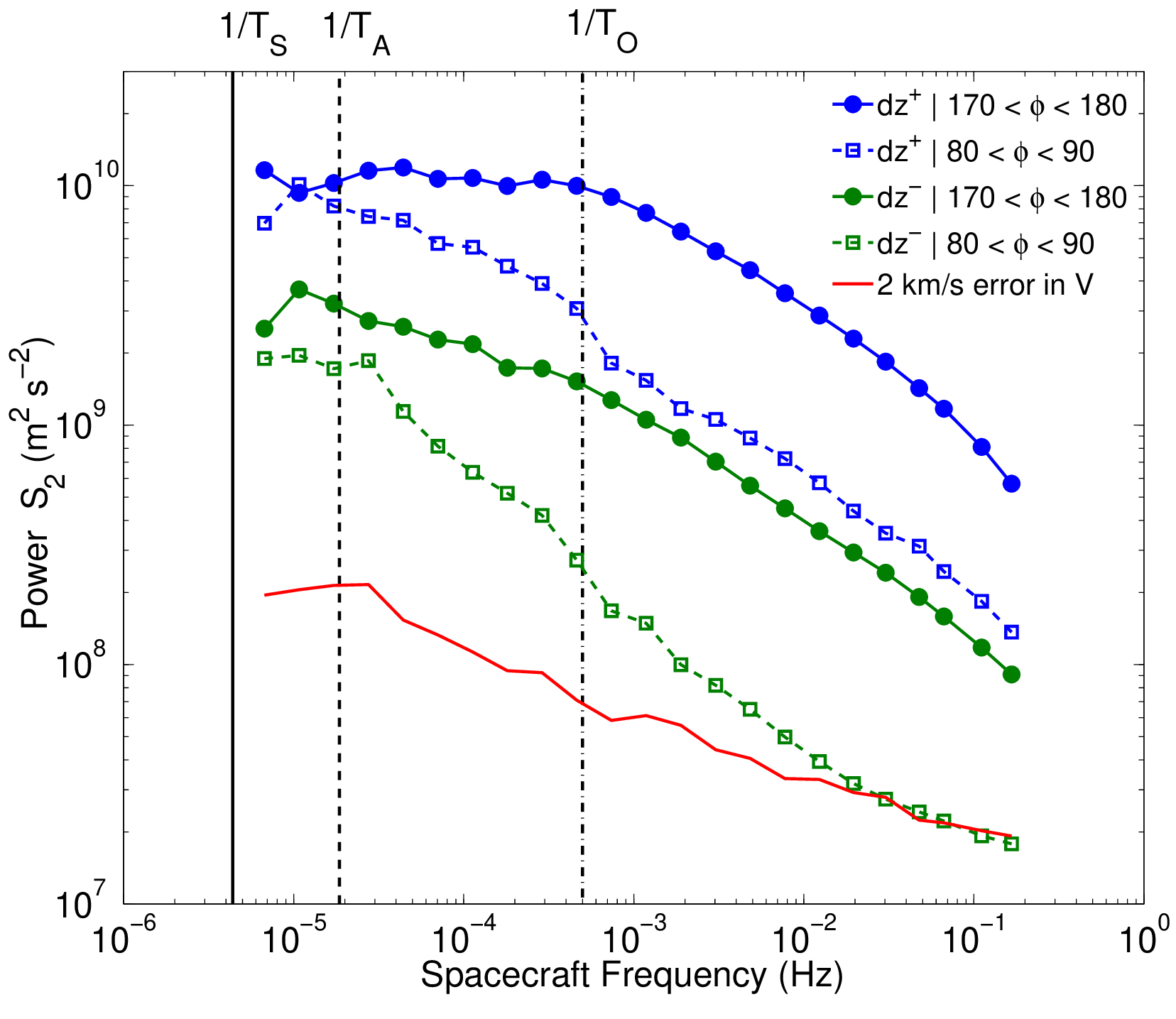}%
\caption{Conditioned structure functions of Elsasser variables from observations of fast solar wind. Anti-aligned (circles) and un-aligned (squares) structure functions of $\delta\textbf{z}^+_{\perp}$ (blue) and $\delta\textbf{z}^-_{\perp}$ (green) are plotted against frequency. Un-aligned fluctuations begin scaling at $f\sim1/T_A$ and have similar gradients to those found in the inertial range, indicating that they may be interacting non-linearly in the $1/f$ range.}
\label{fig:3}
\end{figure}
{\it Relating alignment and spectral scaling.} To investigate whether $\phi$ has any effect on the turbulence and what may be causing the change in $\phi$ with scale, we use structure functions conditioned on $\phi$. Eq.(\ref{eq:S2}) is modified to average over instances when $\phi$ is within a fixed range:
\begin{align}
S_2(\delta\textbf{z}^{\pm}, \tau | \phi_1 < \phi < \phi_2) &= \frac{1}{N}\sum\limits_{\phi(t, \tau) = \phi_1}^{\phi(t, \tau) = \phi_2}{\left| \delta\textbf{z}^{\pm}_{\perp}(t,\tau) \right|^2},\label{eq:S2phi}
\end{align}
where $N$ is the number of points with $\phi_1<\phi<\phi_2$. 
\par
The structure functions calculated according to Eq.~(\ref{eq:S2phi}) are shown in Fig.~\ref{fig:3} for the anti-aligned and perpendicularly aligned fluctuations: $S_2(\delta\textbf{z}^{\pm}, \tau | 170^{\circ}<\phi<180^{\circ})$ (filled symbols) and $S_2(\delta\textbf{z}^{\pm}, \tau | 80^{\circ}<\phi<90^{\circ})$ (open symbols), respectively. This allows us to investigate the effect that the alignment angle has on the turbulence by separating differently aligned fluctuations and observing their scaling in the $1/f$ and inertial ranges.  
\par
\begin{table}%
\begin{tabular}{@{}c@{\ \ }c@{\ \ }cc@{\ \ }c@{}} \hline
&\multicolumn{2}{@{}c@{}}{$1/T_A < f < 1/T_O$}&\multicolumn{2}{@{}c@{}}{$1/T_O < f < 5\times10^{-3}$}\\
&$\!80<\!\phi\!<90$&$\!170<\!\phi\!<180$&$\!80<\!\phi\!<90$&$\!170<\!\phi\!<180$\\ \hline
$\delta\textbf{z}^{+}_{\perp}$&$\!-0.32\pm0.02$&$\!-0.07\pm0.02$&$\!-0.49\pm0.06$&$\!-0.36\pm0.02$\\
$\delta\textbf{z}^{-}_{\perp}$&$\!-0.65\pm0.06$&$\!-0.22\pm0.03$&$\!-0.60\pm0.05$&$\!-0.43\pm0.02$\\[0.2em] \hline
\end{tabular}
\caption{Scaling of the structure functions $S_2(\delta\textbf{z}^{\pm}, \tau | \phi) \propto f^{\alpha}$ of the Elsasser fluctuations in two frequency ranges covering the $1/f$ range and the inertial range. The values are calculated from a linear least squares fit of a straight line to the structure functions on a log-log plot.}
\label{table}
\end{table}
Between $f\sim 1/T_A$ and $f\sim 1/T_O$, the anti-aligned anti-sunward $\delta\textbf{z}^+_{\perp}$ structure functions scale with $\alpha = -0.07\pm0.02$, giving a spectral index close to $-1$. These fluctuations are the most common (Fig.~\ref{fig:2}) and contain the most power and hence dominate the bulk average structure functions in Fig.~\ref{fig:1} - and presumably all previously reported, unconditioned structure functions and spectra in this frequency range of fast solar wind. Perpendicularly aligned $\delta\textbf{z}^+_{\perp}$, however, have a steeper scaling $\alpha = -0.32\pm0.02$, corresponding to a spectral index only slightly shallower than that at higher frequencies in the inertial range, $\alpha = -0.49\pm0.06$. 
\par
The structure functions of perpendicularly aligned sunward fluctuations $\delta\textbf{z}^-_{\perp}$ are steep from $f\sim1/T_A$ until the instrument noise floor (the solid red line; see discussion below) is reached, with $\alpha = -0.65\pm0.06$ in the $1/f$ range and $\alpha = -0.60\pm0.05$ in the inertial range, giving a spectral index close to $-5/3$ in both frequency ranges. The anti-aligned $\delta\textbf{z}^-_{\perp}$ structure functions are flatter in the range $1/T_A < f < 1/T_O$ than in the inertial range, with $\alpha = -0.22\pm0.03$ and $\alpha = -0.43\pm0.02$ respectively. These scalings are summarized in Table~\ref{table}. 
\par
{\it Accuracy of measurements.} Measurement noise is a potential concern in this analysis. The 3DP instrument is known to have noise in the high-cadence moments \cite{Podesta09b} with observations often appearing discretized. By differencing the raw velocity data and finding the most common value we estimate the noise amplitude during the periods we analyze to be approximately equivalent to a 2 km/s uncertainty in each component of {\bf V}. A standard error analysis on Equation (\ref{eq:S2}) leads to the frequency dependent noise represented by the red line in Fig.~\ref{fig:3}. This noise affects the structure functions of perpendicularly aligned sunward fluctuations the most, with the signal to noise ratio of un-aligned $\delta\textbf{z}^{-}_{\perp}$ structure functions becoming significant ($\sim 2$) at $f \sim 2\times10^{-3}$~Hz; this is our estimate of the frequency at which noise begins to render our results unreliable. We have, therefore, restricted the fitting of the structure functions scaling in the inertial range (Table~\ref{table}) to the lowest frequency decade, $1/T_O < f < 5\times10^{-3}$ Hz. This minimizes the effect of the noise on the scaling and we do not draw conclusions about the inertial range.
\par
The noise will also affect the measurement of $\phi$ since it uses the values of $\delta\textbf{z}^{-}_{\perp}$ and $\delta\textbf{z}^{+}_{\perp}$, which in turn contain $\textbf{V}$ observations. The roll-over of $\left<\phi\right>$ at $f \sim 1/T_O$ in Fig.~\ref{fig:2} occurs at a frequency a factor of 4 lower than the frequency $f \sim 2\times10^{-3}$~Hz, at which the signal to noise ratio of the weakest structure functions reaches a value of 2, and so cannot solely be attributed to noise. The strong decrease in $\left<\phi\right>$ and the flattening of the distribution at spacecraft frequencies above $2\times10^{-3}$ Hz is, however, likely to be caused by the noise. 
\par
\begin{figure}
\includegraphics[width=\columnwidth]{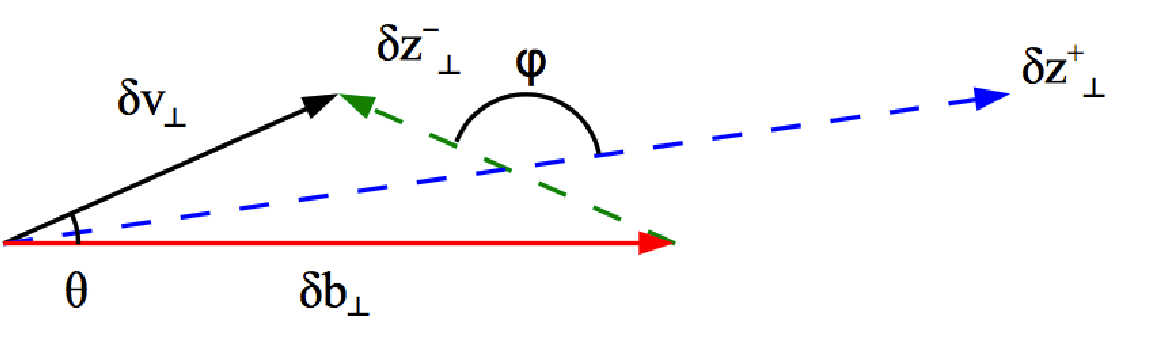}
\caption{Fluctuation vectors and alignment angles in the plane perpendicular to {\bf B}.}
\label{fig:Triangles}
\end{figure}

{\it Alignment and geometry.} Alignment of $\delta\textbf{b}_{\perp}$ and $\delta\textbf{v}_{\perp}$ is related to the alignment of $\delta\textbf{z}^+_{\perp}$ and $\delta\textbf{z}^-_{\perp}$. Fig.~\ref{fig:Triangles} shows a typical geometry in the plane perpendicular to $\textbf{B}_{0}$ in the fast solar wind, assuming $\delta\textbf{b}_{\perp} > \delta\textbf{v}_{\perp}$ with only a small angle $\theta$ between the vectors, as suggested by Fig.~\ref{fig:1} and \ref{fig:2}. This results in $\delta\textbf{z}^+_{\perp} > \delta\textbf{z}^-_{\perp}$ with a large angle $\phi$ between them (anti-alignment). This simple geometry can be expressed in terms of the scale-dependent dimensionless parameters \cite{footnote} normalized cross helicity $\sigma_c = (|\delta\textbf{z}^+_{\perp}|^2-|\delta\textbf{z}^-_{\perp}|^2)/(|\delta\textbf{z}^+_{\perp}|^2+|\delta\textbf{z}^-_{\perp}|^2)$ and normalized residual energy $\sigma_r = (|\delta\textbf{v}_{\perp}|^2-|\delta\textbf{b}_{\perp}|^2)/(|\delta\textbf{v}_{\perp}|^2+|\delta\textbf{b}_{\perp}|^2)$:
\begin{alignat}{2}
\cos(\phi) &= \frac{\delta \textbf{z}^+_{\perp} \cdot \delta \textbf{z}^-_{\perp}}{|\delta \textbf{z}^+_{\perp}||\delta \textbf{z}^-_{\perp}|} &=  \frac{\sigma_r}{\sqrt{1 - \sigma_c^2}}, \label{eq:phidef}\\
\cos(\theta) &= \frac{\delta \textbf{v}_{\perp} \cdot \delta \textbf{b}_{\perp}}{|\delta \textbf{v}_{\perp}||\delta \textbf{b}_{\perp}|} &=  \frac{\sigma_c}{\sqrt{1 - \sigma_r^2}}. \label{eq:thetadef}
\end{alignat}
The geometry in Fig.~\ref{fig:Triangles} is fixed by setting any two of $\theta$, $\phi$, $\sigma_c$ and $\sigma_r$. Therefore, statements about alignment are also statements about normalized cross helicity and residual energy of individual fluctuations, and vice-versa. The fluctuations that scale steeply in Fig.~\ref{fig:3} are those with $\phi \sim 90^{\circ}$ and therefore $\sigma_r \sim 0$ meaning that $\delta\textbf{v}_{\perp}^2 \sim \delta\textbf{b}_{\perp}^2$. 
\par
While we note that the relations in Eq.~(\ref{eq:phidef}, \ref{eq:thetadef}) are only strictly true for individual realizations it is interesting to look at our results concerning averages from this point of view \cite{footnote2}. In the range $1/T_A  \lesssim  f  \lesssim 1/T_O$, $\sigma_r$ is negative and decreases towards $-1$ while $\sigma_c$ increases towards $1$ (Fig.~\ref{fig:1}). So, in accordance with Eq.~(\ref{eq:phidef}), the alignment angle $\phi$ tends towards $180^{\circ}$ (Fig.~\ref{fig:2}). Note that this situation is perhaps consistent with the idea that MHD turbulence would generate negative residual energy \cite{Boldyrev09}. 
In the same vein, we conclude from Eq.~(\ref{eq:thetadef}) that the alignment between $\delta\textbf{v}_{\perp}$ and $\delta\textbf{b}_{\perp}$ intensifies in the $1/f$ range, as indeed seen in the solar wind (the angle $\theta$ is shown in white in Fig.~\ref{fig:2} and was previously measured in \cite{Podesta09b, Hnat11}).
\par
{\it Discussion.} We have shown that in the fast solar wind, both the distribution and the mean of the angle between Elsasser fluctuations, $\phi$ are scale dependent. The probability of Elsasser fluctuations being anti-aligned ($\phi \sim 180^{\circ}$) starts increasing at the scale at which Alfv\'{e}n waves begin to interact, $f\sim1/T_A$, and stops at the outer scale, $f\sim1/T_O$. 
\par
In Fig.~\ref{fig:3} and the attendant discussion, we showed that the $1/f$ scaling of spectra (flat scaling of structure functions) detected in numerous previous observations of fast solar wind is dominated by the anti-aligned subset of the anti-sunward Elsasser fluctuations ($\delta\textbf{z}^+_{\perp}$). Hidden beneath this energetically dominant sea of non-interacting (or weakly interacting) `non-turbulent' fluctuations are the unaligned fluctuations, which exhibit steep spectral scalings symptomatic of a nonlinear cascade. We hypothesize that the steep scaling of unaligned fluctuations is caused by the increased {\it in-situ} non-linear interaction of these fluctuations, since both populations of fluctuations have similar $\sigma_c$ and travel time from the Sun. The different behavior of these populations is reminiscent of the difference between slow and fast solar wind streams \cite{Grappin}, however the fluctuations that scale steeply ($\sigma_r = 0$) do not resemble those characteristic of slow wind ($\sigma_r < 0$). 
\par
It is an interesting question whether the anti-aligned `non-turbulent' fluctuations are Alfv\'{e}n waves or magnetically dominated force-free structures. The pure case of the former would require $\delta\textbf{z}^+_{\perp} \gg \delta\textbf{z}^-_{\perp}$ and so $\delta\textbf{b}_{\perp} \sim \delta\textbf{v}_{\perp}$ ($\sigma_c \approx 1$, $|\sigma_r| \ll 1$); the pure case of the latter, $\delta\textbf{b}_{\perp} \gg \delta\textbf{v}_{\perp}$ and so $\delta\textbf{z}^+_{\perp} \sim \delta\textbf{z}^-_{\perp}$ ($\sigma_r \approx 1$, $|\sigma_c| \ll 1$). The measured fluctuations appear to be in between these two extremes ($\delta\textbf{b}_{\perp} > \delta\textbf{v}_{\perp}$ and $\delta\textbf{z}^+_{\perp} > \delta\textbf{z}^-_{\perp}$) and can perhaps be interpreted as a mixture of them \cite{Tu91, Tu93}. Both types of fluctuation are slow to decay; this can be thought of in terms of conservation of cross helicity (Alfv\'{e}n waves) and magnetic helicity (force-free structures, subject to the minimum-energy constant-helicity relaxation principle \cite{Taylor74, Woltjer58}). The generation of residual energy at low frequencies \cite{Boldyrev09} could then be interpreted as generation (or occurrence and persistence) of force-free structures.
\par
We conclude that the turbulent cascade in the fast solar wind starts at larger scales than previously thought, although it is restricted to perpendicularly aligned fluctuations and energetically sub-dominant. Measured scale-dependent alignment in the $1/f$ range represents the change in the fractional populations with scale of turbulent, non-linearly interacting, perpendicularly aligned fluctuations versus non-interacting, anti-aligned fluctuations. We have identified a new, larger, outer scale ($f\sim1/T_A$), which is consistent with an anti-sunward Alfv\'{e}n wave requiring only one interaction with an oppositely directed wave to launch the perpendicularly aligned cascade. It is still uncertain what determines the frequency at which the spectral break between the $1/f$ and the inertial ranges occurs \cite{Matthaeus86, Bavassano92, Matthaeus07, Verdini12, Grappin}.
\par
This research was supported by the NASA Postdoctoral Program at the Goddard Space Flight Center (RTW), STFC (RTW, AM, TSH), NASA grant NNX09AE41G (CHKC) and the Leverhulme Trust Network for Magnetized Plasma Turbulence. Wind data were obtained from the NSSDC website http://nssdc.gsfc.nasa.gov.


\begin{thebibliography}{99}
\bibitem[Footnote(2012)]{footnote}
Note that these relations have been used in previous studies of the character of solar wind turbulence \cite{Bavassano98, Bavassano06} where they are called `correlation functions'. There has also been a more recent study \citep{Osman11} which shows that the alignment angle $\theta$ depends directly on cross helicity, which can be clearly understood from Eq.~(\ref{eq:thetadef}).
\bibitem[Footnote2(2012)]{footnote2}
General inferences can be made for the mean values because observations of the PDFs of $\phi$ and for $\theta$, $\sigma_c$ and $\sigma_r$ \citep{Podesta09b, Osman11, Bavassano98, Bavassano06} are singly peaked and the means follow the general trend of the distributions with scale.
\end{thebibliography}
\end{document}